\pdfoutput=1
\documentclass[%
 reprint,
 amsmath,amssymb,
 aps,
]{revtex4-2}

\usepackage{graphicx}
\usepackage{dcolumn}
\usepackage{bm}
\usepackage{xcolor}
\usepackage{multirow}

\begin{document}

\preprint{APS/123-QED}

\title{Ordinary muon capture rates on $^{100}$Mo and $^{\rm nat}$Mo for astro-antineutrinos and double beta decays}

\author{I. H. Hashim}
\email[corresponding author: ]{izyan@utm.my}
\altaffiliation[Also at ]{National Centre of Particle Physics, Universiti Malaya, 50603 Kuala Lumpur, Malaysia.}
\altaffiliation[Also at ]{UTM-Centre of Industrial and Applied Mathematics, Universiti Teknologi Malaysia, 81310 Johor Bahru, Johor, Malaysia.Malaysia.}
\author{N. N. A. M. A. Ghani}
\author{F. Othman}
\author{R. Razali}
\author{Z. W. Ng}
\affiliation{%
Department of Physics, Faculty of Science, Universiti Teknologi Malaysia, 81310 Johor Bahru, Johor, Malaysia.}%


\author{H. Ejiri}
\author{T. Shima}
\author{D. Tomono}
\affiliation{Research Centre for Nuclear Physics, 10-1 Mihogaoka, Ibaraki, Osaka 567-0047, Japan.}%

\author{D. Zinatulina}
\author{M. Schirchenko}
\author{S. Kazartsev}
\affiliation{Joint Institute for Nuclear Research, 6 Joliot-Curie St, 141980 Dubna, Moscow Region, Russia}

\author{A. Sato}
\author{Y. Kawashima}
\affiliation{Department of Physics, Graduate School of Science, Osaka University, Machikaneyama 1-1, Toyonaka, Osaka 564-0043, Japan}

\author{K. Ninomiya}%
\affiliation{Department of Chemistry, Graduate School of Science, Osaka University, Machikaneyama 1-1, Toyonaka, Osaka 564-0043, Japan}

\author{K. Takahisa}
\affiliation{Department of Radiological Technology, Faculty of Health Sciences, Kobe Tokiwa University, Otani 2-6-2, Nagata, Kobe, Hyogo 653-0838, Japan}

\date{\today}

\begin{abstract}
\begin{description}
\item[Background] The nuclear responses for antineutrinos associated with double beta decays (DBDs) and astro-antineutrino interactions are studied by measuring ordinary muon capture (OMC) rates.
\item[Purpose]The experimental studies of absolute OMC rates and their mass number dependence for $^{100}$Mo and the natural Mo are currently of interest in astro-antineutrinos and DBDs.
\item[Method]The OMC rates were obtained experimentally by measuring the time spectrum of the trapped muon's decay into electrons to obtain the half-lives of the trapped muons.
\item[Results]The OMC rate for the enriched isotope of $^{100}$Mo is $\Lambda$($^{100}$Mo)=(7.07$\pm$0.32)$\times10^{6}$ s$^{-1}$, while that for the natural Mo is $\Lambda$($^{\rm nat}$Mo)=(9.66$\pm$0.44)$\times10^{6}$ s$^{-1}$, i.e., $\Lambda$($^{100}$Mo) is about 27$\%$ of $\Lambda$($^{\rm nat}$Mo), reflecting the blocking effect of the excess neutrons for the proton-to-neutron transformation in OMC. 
The present experimental observation is consistent with the predictions using Goulard-Primakoff's (GPs) and Primakoff's (Ps) empirical equations. 
\item[Conclusions] The absolute OMC rates for $^{100}$Mo and $^{\rm nat}$Mo were measured. The large neutron excess in $^{100}$Mo gives a much lower OMC rate than $^{\rm nat}$Mo. 
On both $^{100}$Mo and $^{\rm nat}$Mo, consistent OMC rates with the GP and P values are observed.  
\end{description}
\end{abstract}

\maketitle


\section{Introduction}

Neutrinoless double beta (0$\nu\beta\beta$) decays and supernova neutrinos are of current interest for neutrino studies beyond and within the standard model. 
They are crucial for studying neutrino properties and weak interactions beyond the electro-weak standard theory and for investigating supernova neutrino/antineutrino nuclear syntheses and nuclear interactions, as addressed in the review papers \cite{Ejiri2019PR,Vergados2012RRP, Ejiri2021Fphys,Ejiri2002PLB} and references therein. 
To acquire 0$\nu\beta\beta$ decay rates ($R^{0\nu}$), we must first calculate the coherent sum of the individual matrix elements $M^{0\nu}_i$ \cite{Hashim2021Fspas}. 
For each intermediate state $i$, there are two corresponding isospin $\tau^{\pm}$ directions that are now being investigated using nuclear, lepton, and photon probes \cite{Ejiri2019PR,Ejiri2015JPhysG}.

Neutrino responses for $\tau^-$ have been studied by charge exchange reactions (CER) using ($^3$He,$t$) to evaluate the nuclear matrix element, $M$($\beta^-$). 
The $\tau^+$ responses, on the other hand, are studied by OMC reactions and photonuclear reactions via isobaric analogue states (IAS) \cite{Ejiri2019PR,Ejiri2021Fphys}.
OMC is a weak nuclear process via the exchange of a charged weak boson and the negative muon capture on a medium-heavy nucleus $^{A}_{Z}$X forms a highly excited compound nucleus of $^A_{Z-1}$Y with the excitation energy up to the muon mass of 100 MeV.
In the case of OMC, a muon neutrino is emitted, taking around 100-50 MeV of energy. 
The OMC leaves the residual nucleus $^{A}_{Z-1}$Y in the range of excitation energy $E$ = 0-50 MeV and momentum transfer of 100-50 MeV c$^{-1}$, analogous to DBDs and supernova neutrino interactions.

The $\tau^+$ responses in medium-heavy nuclei involving proton ($p$) to neutron ($n$) transformation are sensitive to the neutron excess due to the generation of additional neutrons.
It is worth noting that most $\beta^-\beta^-$ decays are studied by using isotopes with the highest $N-Z$.
Experimental studies show that the OMC rate decreases with the mass number $A$ \cite{Suzuki1987PRC,Sens1959PR,Daniya2019PRC,Fynbo2003NPA,Mamedov2000JETPL,Measday2001PR}.
Nuclei with larger $N$ have a much lower OMC rate due to the blocking of low-energy $p$-to-$n$ transitions by the excess neutrons in the nuclei.
A significant increment in the OMC rate is observed when the second and higher forbidden transitions are involved \cite{Fynbo2003NPA}.

The present work aims to study the OMC rates for $^{100}$Mo and $^{\rm nat}$Mo to investigate the $N-Z$ dependence of the OMC on Mo isotopes, which are of interest for the astro-antineutrinos \cite{Ejiri2002PLB} and DBD \cite{Ejiri2019PR}. 
The OMC rate of $\Lambda_{\mu}^{\rm cap}$ associated with the $\tau^{+}$ response can be evaluated directly through the muon disappearance rate by observing the electrons from free muon decays as a function of time \cite{Daniya2019PRC,Measday2001PR}. 
The total muon disappearance rate $\Lambda_{\mu}^{\rm total}$ is given by 
\begin{equation}
\Lambda_{\mu}^{\rm total} = \frac{1}{{\tau}_{\mu}} = \Lambda_{\mu}^{\rm cap} + H \times \Lambda_{\rm decay}
\label{eq:caprate}
\end{equation}

\noindent where $\Lambda_{\mu}^{\rm cap}$ indicates the OMC rate, $H$ is the Huff factor, and the free muon decay rate is given by $\Lambda_{\rm decay}$=4.5$\times$ 10$^6$ s$^{-1}$. 
In this case, the total rate is equal to the inverse of the muon absolute lifetime, $\tau_{\mu}$.

The review article \cite{Measday2001PR} discusses great details about the OMC rates.
A proton-neutron emission model (PNEM) has been developed to examine the overall distribution of final states following the OMC reaction as well as the branching ratios of residual radioactive isotopes (RIs) following OMC \cite{Hashim2018PRC,Hashim2020NIMA,Hashim2021Fspas}.
In this case, the experimental branching ratios of RIs after OMC are compared with PNEM output to detect the significant giant resonance (GR) peak for the reaction.
The universal axial vector reductions from a single beta matrix $M(\beta^-)$ from CER on $^{100}$Mo have so far been published in \cite{Thies2012PRC}.

In the present work, we report, for the first time, experimental studies of the OMC rate for $^{100}$Mo, which is of interest for DBD and astro-antineutrino, and the $A$ or $N-Z$ dependence of the rates by comparing it with the $^{\rm nat}$Mo rate.
A comparison with the GP and P empirical rates and trends from experimental OMC rates are also discussed.

\section{Methodology}

The $^{100}$Mo target is one of the candidates for supernova neutrinos and supernova antineutrinos detections \cite{Ejiri2000PRL,Ejiri2019PR,Simkovic2020PRC,Barabash2019Fphys,cremonsi2014AHEP,Ejiri2002PLB,Stephan2019AAS}.
$^{100}$Mo is an interesting nucleus because of the large neutron excess (note that since nuclei used for DBD are mostly those with the largest $N-Z$).

As the targets, a 255.1 mg cm$^{-2}$ thick $^{100}$Mo foil with 96.5$\%$ enrichment and a 4.1 g cm$^{-2}$ thick $\rm ^{nat}$Mo slab were utilized.
A 3 mm thick Al degrader was used in the $^{100}$Mo irradiation to slow down the incoming muons and stop them in the targets.
The degrader was removed during the $^{\rm nat}$Mo irradiation.
The experiment was carried out at Osaka University's Research Center for Nuclear Physics (RCNP) using a 400 MeV proton beam accelerated by the ring cyclotron.
Muons are produced at the MuSIC facility through proton-graphite interactions.
The 45-55 MeV c$^{-1}$ muon beam ($\approx$ 10$^4$-10$^5$ muons s$^{-1}$) was focused on a target 5 cm away from the beam exit.
Each target took about eight hours to run.

\begin{figure}[htb!]
\includegraphics[scale = 0.5]{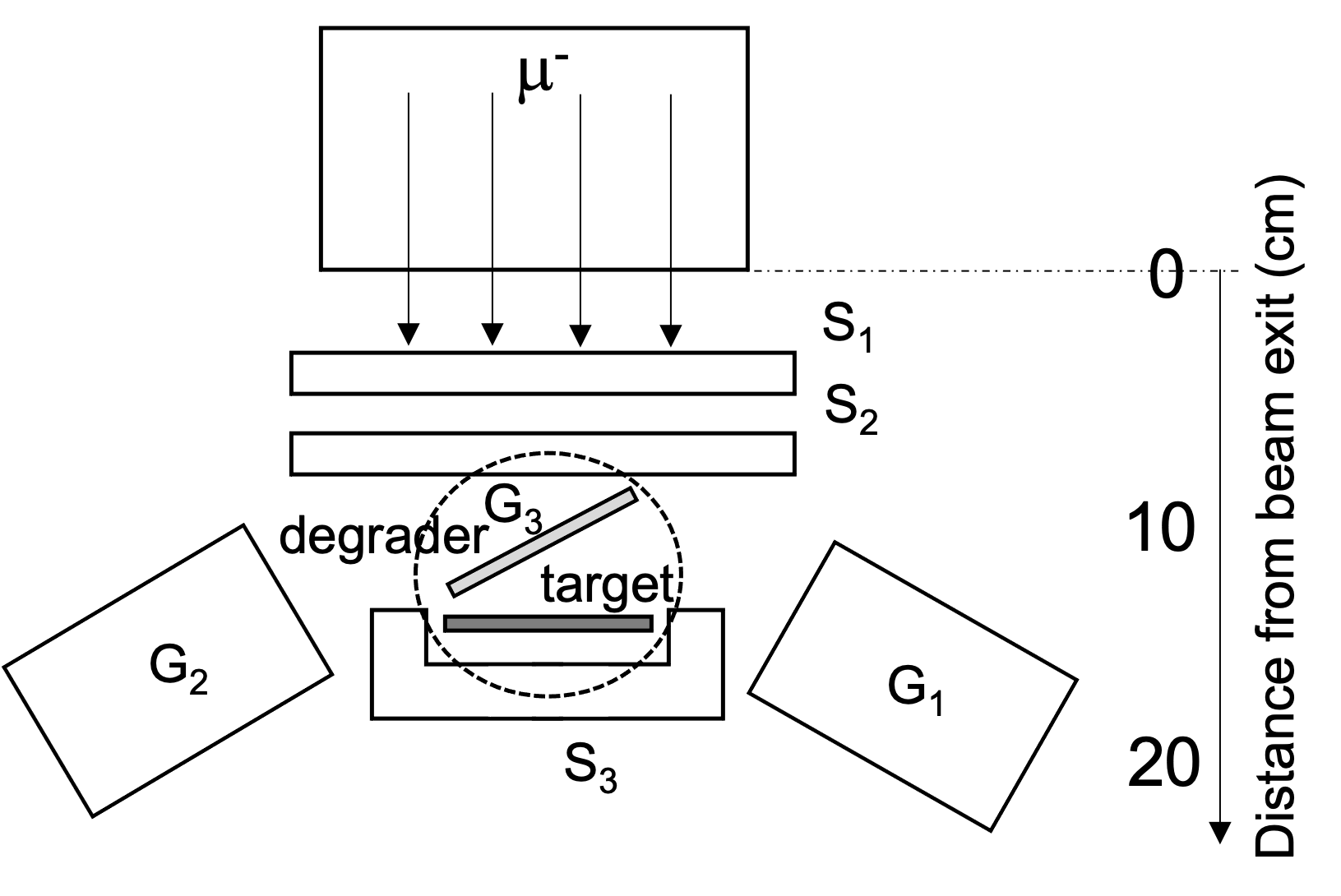}
\caption{\label{fig:1} Muon irradiation on $^{\rm nat}$Mo and $^{100}$Mo at RCNP, Osaka University's MuSIC facility. The scintillation counters are S$_1$, S$_2$, and S$_3$, while the HPGe detectors are G$_1$, G$_2$, and G$_3$. The data for the absolute muon lifetime were obtained using the scintillation detectors to get the stop signal under the conditions of signals from S$_1$ and S$_2$ and no signal for S$_3$.}
\end{figure}

The experimental setup is depicted in Fig. \ref{fig:1}.
The negative muon triggering two scintillation counters (S$_1$ and S$_2$) and no signals from the counter (S$_3$) indicate that the muon is stopped at the Mo target.
Outgoing gamma rays and muonic X-rays were immediately recorded using one planar type and two coaxial end types of HPGe detectors (designated as G$_1$, G$_2$, and G$_3$).
The muonic X-ray signal was recorded using three ADC channels (ch0, ch1, and ch2).
Other channels (ch3, ch4, and ch5) were allocated to gamma rays with a short lifetime.
Within a 1 $\mu$s time window, the time spectrum of the free-decay electron from muon stopping events on the target is recorded in ch10.

\section{Results and Discussion}

\begin{figure}[htb!]
\includegraphics[scale = 0.35]{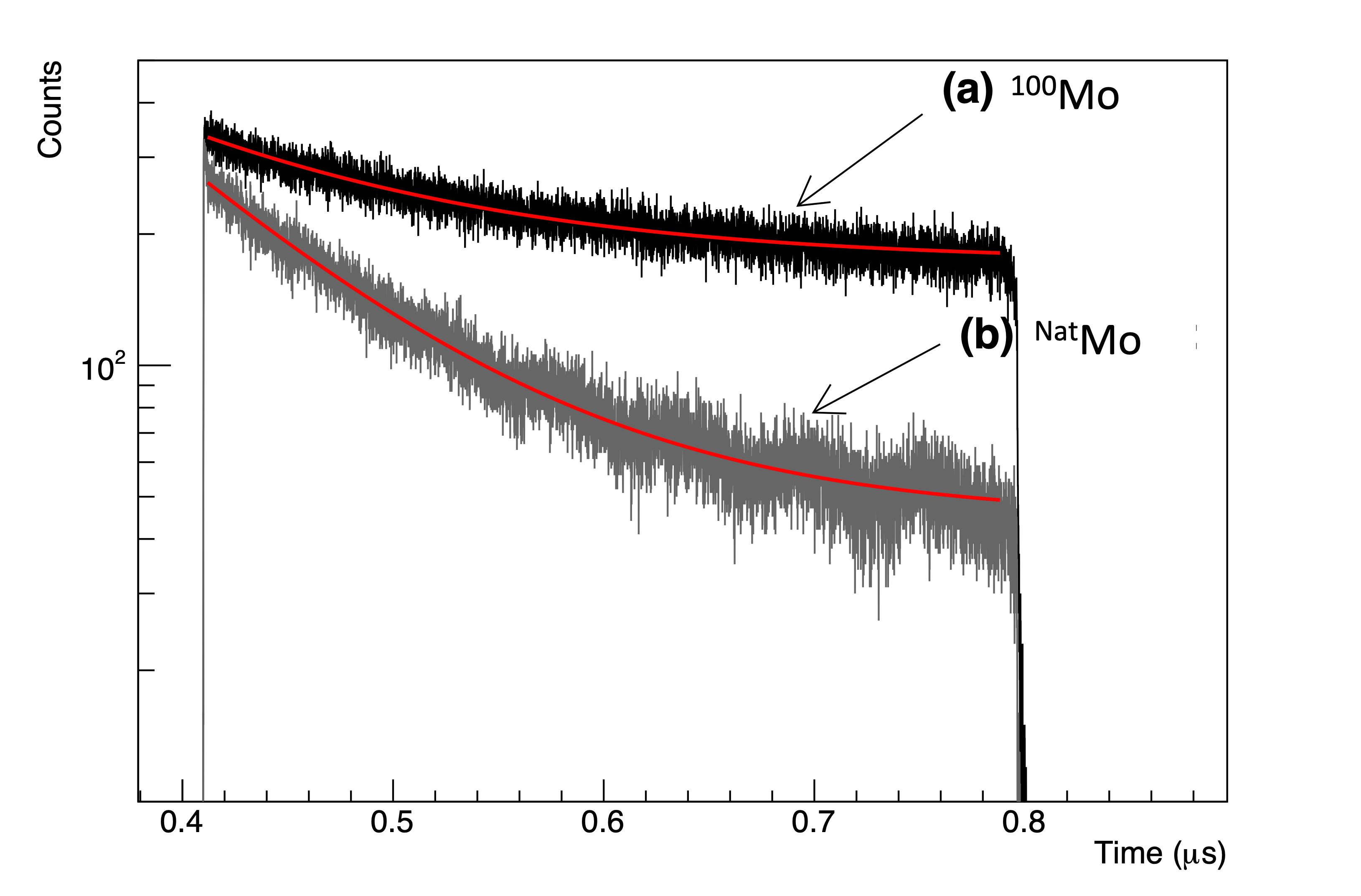}
\caption{\label{fig:2} Electron time distributions for (a) $^{100}$Mo and (b) $^{\rm nat}$Mo.}
\end{figure}

Muons are stopped at the target and trapped in the atom's inner orbit and then disappear either by decaying into a muon neutrino and an electron, or by muon capture on the nucleus.
The total rate in eq. (\ref{eq:caprate}), i.e., the inverse of the muon lifetime, was measured by observing the electron from the free muon decays as a function of time. 
The electron time distributions for enriched $^{100}$Mo and $^{\rm nat}$Mo were shown in Fig. \ref{fig:2}, and the fitting results were listed in Table \ref{tab:table1}. 

\begin{table}[htb!]
\caption{\label{tab:table1}Fit results of the $^{100}$Mo and $^{\rm nat}$Mo (in ch10, 100 MHz TDC). $\Delta$ refers to the fitting error.}
\begin{ruledtabular}
\begin{tabular}{ccccccc}
Isotope & $\tau_{\mu}$ (ns) & $\Delta\tau_{\mu}$ (ns) & $N_{0}$ & $\Delta N_{0}$ & $N_{BG}$ & $\Delta N_{BG}$ \\ \hline
$^{100}$Mo & 133.4 & 6.0 & 3568 & 142.7 & 170.2 & 6.8 \\
$^{\rm nat}$Mo & 99.1 & 4.5 & 13924 & 542.7 & 43.3 & 0.8 \\
\end{tabular}
\end{ruledtabular}
\end{table}

The muon absolute lifetimes for $^{100}$Mo and $^{\rm nat}$Mo were evaluated from the observed time spectra of the electrons within a 1 $\mu$s time window from the muon stop. 
The measured decay curves were fitted by $N(t)$=$N_{BG}$+$N_0(t)$ $\times$ exp(-$t/\tau_{\mu}$) assuming that the hyperfine effect on molybdenum is very small \cite{Measday2001PR}. 
Here $N_{BG}$ represents the background counts, $N_0(t)$ represents the number of initial isotopes generated at time $t$, $t$ represents the measurement time, and $\tau_{\mu}$ represents the muon absolute lifetime.
The time distributions for $^{100}$Mo and $^{\rm nat}$Mo show a simple exponential decay curve with a large statistical fluctuation throughout the measurement due to the short measurement time.
The fit results in Table \ref{tab:table1} include a 4$\%$ systematic error from detector efficiency and the gain shift during non-stop monitoring with low statistics.

The absolute lifetimes for muons in $^{100}$Mo and $^{\rm nat}$Mo are $\tau_{\mu}$($^{100}$Mo)=133.4$\pm$6.0 ns and $\tau_{\mu}$($^{\rm nat}$Mo)=99.1$\pm$4.5 ns. 
$\tau_{\mu}$($^{100}$Mo) is greater than $\tau_{\mu}$($^{\rm nat}$Mo) by a factor of 1.37.
The OMC rates for $^{100}$Mo and $^{\rm nat}$Mo are evaluated using eq. (\ref{eq:caprate}) with the measured value of $\mu$ lifetimes presented in Table \ref{tab:table1}. 
The OMC rates for $^{\rm nat}$Mo and $^{100}$Mo are derived as (9.66$\pm$0.44)$\times$10$^6$ s$^{-1}$ and (7.07$\pm$0.32)$\times$10$^6$ s$^{-1}$ respectively, as shown in Table \ref{tab:table2}.
The decrement of the OMC rate for different $A$ shows the $N-Z$ dependence.

\begin{table}[htb!]
\caption{\label{tab:table2} Comparison of OMC rates for Mo nuclei reported by \cite{Suzuki1987PRC} (in column 2), and the present experiment in column 3.}
\begin{ruledtabular}
\begin{tabular}{ccc}
Isotopes & $\Lambda^{\rm cap}_{\mu} \times 10^{6}$ s$^{-1}$ \cite{Suzuki1987PRC} & $\Lambda^{\rm cap}_{\mu} \times 10^{6}$ s$^{-1}$ (this work)   \\ \hline
{$^{\rm nat}$Mo} & 9.614 $\pm$ 0.15 & 9.66 $\pm$ 0.44  \\ 
$^{100}$Mo      &   &  7.07 $\pm$ 0.32   \\
\end{tabular}
\end{ruledtabular}
\end{table}

The present value of 9.66 $\times$ 10$^{6}$ s$^{-1}$ agrees with the value of 9.614 $\times$ 10$^{6}$ s$^{-1}$ by \cite{Suzuki1987PRC} for $^{\rm nat}$Mo within the errors.
Note that \cite{Suzuki1987PRC} uses $^{\rm nat}$Mo with a weighted mass average of $A$=96.
On the other hand, the OMC rate in this study for $^{100}$Mo is smaller than $^{\rm nat}$Mo.
The OMC rate by Goulard Primakoff (GP) is given by eq.(\ref{eq:GPeq})

\begin{eqnarray}
\Lambda_{\mu}^{GP}(A,Z)=Z_{\rm eff}^4 G_1 \left( 1+G_2\frac{A}{2Z}-G_3\frac{A-2Z}{2Z} \right) \nonumber\\
- Z_{\rm eff}^4 G_1 G_4 \left( \frac{A-Z}{2A}+\frac{A-2Z}{8AZ}\right)  \nonumber\\
 \label{eq:GPeq}
\end{eqnarray}

\noindent where $G_1$ = 261, $G_2$ = -0.040, $G_3$ = -0.26 and $G_4$ = 3.24. 
Additionally, Primakoff's (P) empirical formula is given by

\begin{eqnarray}
\Lambda_{\mu}^{P}(A,Z) = Z_{\rm eff}^4 X_1 \left( 1-X_2\frac{A-Z}{2A} \right) 
 \label{eq:Peq}
\end{eqnarray}

\noindent where $X_1$ = 170 s$^{-1}$ and $X_2$ = 3.125.
The recommended discrepancy between the GP and P empirical formulas for individual isotopes is set to be around 10$\%$ \cite{Simkovic2020PRC}.

\begin{figure}[htb!]
\includegraphics[scale = 0.35]{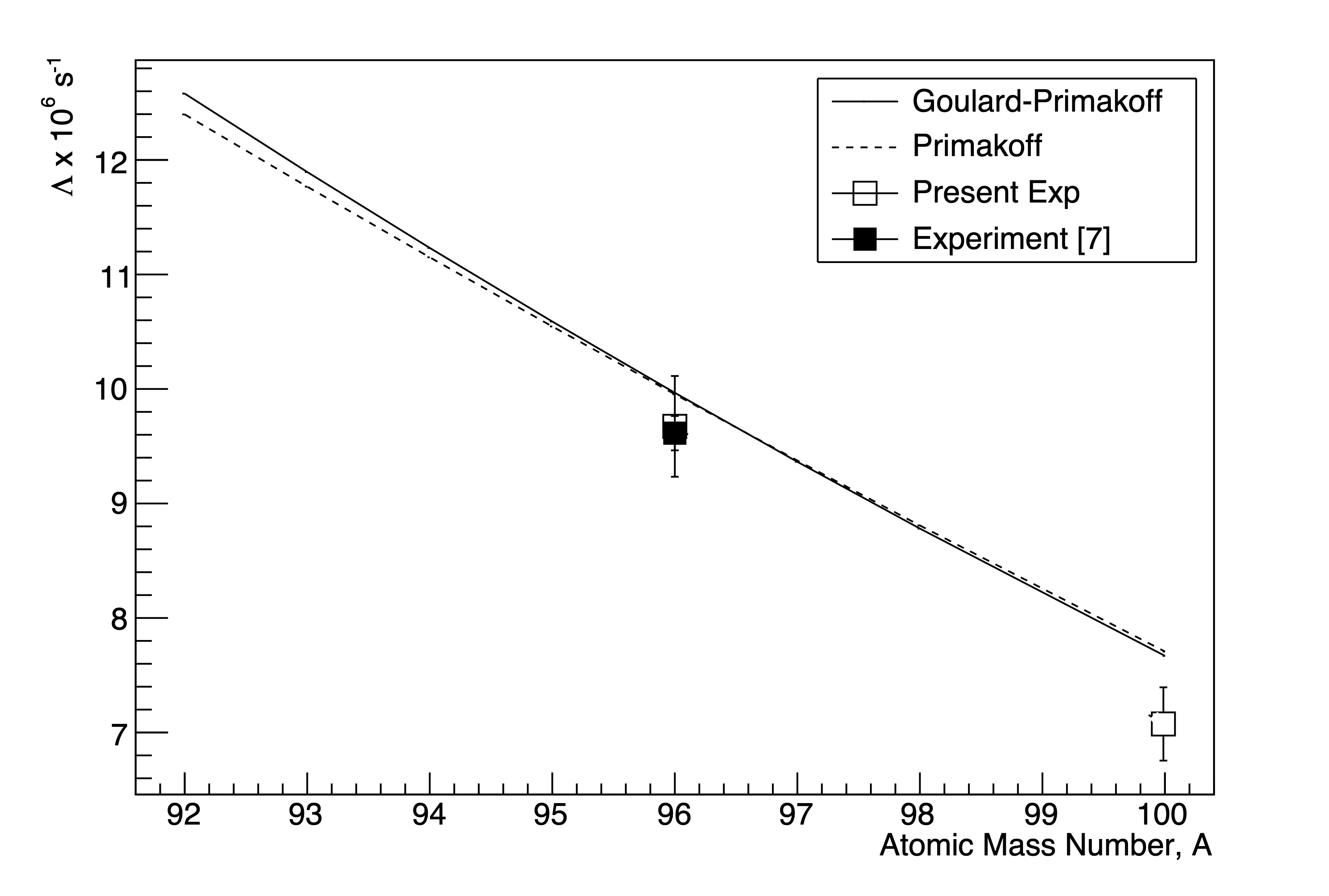}
\caption{\label{fig:3} Calculated and experimental OMC rates as a function of $A$ for all Mo isotopes. 
A small difference is observed in the low $A$ region between GP and P values.}
\end{figure}

Fig. \ref{fig:3} illustrates the overall calculations using the GP and P empirical formulas for molybdenum isotopes with 92 $\leq A \leq$ 100. 
The present experimental data obtained from the above evaluation and the experimental data reported in \cite{Suzuki1987PRC} are included for comparison.
Both empirical equations show decreasing trends for Mo isotopes as the mass number $A$ increases.
The GP and P values are slightly deflected at low $A$ due to the inclusion of higher-order Pauli corrections for heavy elements in the GP empirical formula.

In the present work, the experimental data agree with both the GP and P empirical values. 
The present OMC rates for $^{\rm nat}$Mo and $^{100}$Mo are within 10$\%$ errors of GP and P values, where the OMC rate for $A$ = 100 is 27$\%$ lower than that for $A$ = 96.
The experimental rate for $^{\rm nat}$Mo from \cite{Suzuki1987PRC} yields a much lower value (around 9$\%$) than the GP and P values.
The OMC rate for present $^{\rm nat}$Mo and from \cite{Suzuki1987PRC} is very close to the GP and P empirical values.
However, the current $^{100}$Mo provides a greater departure from the empirical values.
This finding suggests that the empirical GP and P formulations do not properly account for the isotope effect at very large $N-Z$ \cite{Suzuki1987PRC, Measday2001PR}.

DBD nuclei of current interest are $^{76}$Ge, $^{96}$Zr, $^{100}$Mo, $^{116}$Cd, $^{130}$Te, $^{136}$Xe, and $^{150}$Nd.
All of them have the large neutron excess, $\Delta N \approx$ 4-5, more than the average $N$ of the same isotopes (same $Z$).
Thus, their OMC rates and the $\tau^+$ responses are considered to be much reduced due to the neutron excess. 
This has a substantial impact on OMC rates for DBD nuclei.

On a variety of compound and enriched nuclei, OMC rates have been measured using either electron decays or neutron decays \cite{Suzuki1987PRC,Measday2001PR,Daniya2019PRC}.
The electron decays and the neutron decays for OMC rates have shown relatively consistent results. 
References' \cite{Measday2001PR,Daniya2019PRC} reported the present OMC rates vary from 450 to 12.6 $\times$ 10$^6$ s$^{-1}$, with an increment of $A$ resulting in significantly lower OMC rates. 
Further analysis of Ca isotope using random phase approximation (RPA) \cite{Kolbe2001EPJA} shows that the $N$ excess has a considerable impact on the OMC rate.
In this case, the $N$ $\approx$ $Z$ isotopes have a consistently higher OMC rate than the neutron surplus nucleus.
Similarly, neutron excess in a medium-heavy nucleus reduces $\beta^+$ and antineutrino responses by inhibiting 1$^+$ Gamow-Teller (GT) excitations \cite{Measday2001PR,Hashim2018PRC, Zinner2006PRC}. 
Additional research into the $N-Z$ dependence for the strength distributions will be fascinating to better understand the neutrino nuclear responses using OMC processes.

Experimental strength distributions from OMC experiments are currently under development using PNEM for DBD nuclei and available OMC data \cite{Measday2001PR,Suzuki1987PRC,Daniya2019PRC, Measday2007AIP}.
This work was further used for the evaluation of the axial vector coupling constant $g_A$ in comparison to $^{100}$Mo capture strength using pn-QRPA in \cite{Lotta2019PLB}.
Here, the main contribution is due to the low spin states of 0$^{+}$, 1$^{\pm}$ and 2$^{\pm}$ with about 10$\%$ to 15$\%$ coming from higher spin states.
Furthermore, the effective axial vector coupling constant of $g_{A}^{\rm eff}$ $\geq$ 1 has been shown to reproduce well the observed rates for $^{\rm nat}$Mo and $^{100}$Mo in the SRPA and the QRPA \cite{Simkovic2020PRC,Zinner2006PRC}, and pn-QRPA calculations in \cite{Ciccarelli2020PRC}.
Our OMC rates for $^{100}$Mo and $^{\rm nat}$Mo are within 5$\%$ different from the SRPA/QRPA \cite{Zinner2006PRC,Simkovic2020PRC}.
Whereas the pn-QRPA by \cite{Ciccarelli2020PRC}, which predicts well the observed OMC rate, uses the Foldy-Walecka formalism with additional theoretical frameworks to highlight the effect of nuclear structure on weak lepton-nucleus interactions.
On the other hand, the pn-QRPA theoretical calculation using Morita-Fujii formalism \cite{Lotta2019PLB,Lotta2020PRC,Lotta2021Fphys} gives much higher OMC rates.

Based on comparisons with numerous experimental data and the recent theoretical model calculations, the values for the effective axial vector coupling constant $g_{A}^{\rm eff}$ are around 0.5 \cite {Lotta2019PLB,Lotta2019PRC} and 1.0-1.27 \cite {Ciccarelli2020PRC,Simkovic2020PRC}.
The extraction of the $\tau^+$ responses from the muon capture strength could aid in determining the best $g^{\rm eff}_A$ to use in reproducing the absolute capture strength and understanding the $M^{0\nu}$.

\section{Concluding Remarks}

OMC on the nucleus is a weak semileptonic process that can be used to explore neutrino nuclear responses important for DBD and astro-antineutrino nuclear responses.
It can provide information on the $\tau^+$ responses associated with the DBD NME intermediate states.
The $\tau^+$ responses describe the overall final state distribution following the muon capture procedure.

The experimental OMC rates for $^{100}$Mo and $^{\rm nat}$Mo are determined by examining the time spectra of the electron decaying from the trapped muon.
The OMC rate for $^{100}$Mo is 27$\%$ lower than the rate for $^{\rm nat}$Mo, with the effective $A$ around 96 due to the blocking effect of the surplus neutrons in $^{100}$Mo.
The present experimental OMC rates are within 10$\%$ of the GP and P predictions, which are consistent with prior investigations that used enriched nuclei.
This demonstrates that some $g_A$ quenching was detected in nuclear structural effects on the OMC process.
Using this suggested $g_{A}^{\rm eff}$ range, the neutrino nuclear response (square of absolute NME) for DBD nuclei can, on the other hand, be calculated using the pn-QRPA theoretical calculation, which is also being done at Jyvaskyla \cite{Lotta2020PRC,Lotta2021Fphys}.
The absolute and relative OMC rates as a function of excitation energy may aid theories in evaluating 0$\nu\beta\beta$ NMEs and astro-antineutrino synthesis/interaction NMEs.
The ordinary muon capture for double beta decay (OMC4DBD/MONUMENT) collaboration is conducting extensive experimental programs on OMCs to study nuclear responses for DBD neutrinos and supernova neutrinos at the RCNP in Osaka, Japan, and the Paul Scherrer Institute (PSI) in Zurich. 
There is a plan for remeasuring the mass-number distribution of RIs from OMC on $^{100}$Mo at PSI to check the possibility of extracting the partial OMC rates.

\begin{acknowledgments}
The authors thank RCNP, Osaka University for the muon beam used for this experiment.
I.H.Hashim would like to acknowledge the financial support from UTM (R.J130000.7854.5F227) and the Ministry of Higher Education Malaysia (FRGS/1/2019/STG02/UTM/02/6).
I.H. Hashim is grateful to NCPP Management, Universiti Malaya, for hosting the research fellowship.
The reported study was partially funded by Russian Foundation for Basic Research (RFBR) and German Research Foundation (DFG) with the associated project number 21-52-12040.
\end{acknowledgments}

\bibliography{apssamp}

\end{document}